\pdfoutput=1
\documentclass[aps,prl,reprint,nofootinbib,preprintnumbers,amsmath,amssymb,amsfonts]{revtex4-1}

\usepackage[utf8]{inputenc}
\usepackage{graphicx}
\usepackage[breaklinks]{hyperref}

\newcommand{\df}{\mathrel{:=}}

\newcommand{\calF}{\mathcal{F}}

\begin{document}

\title{Infinite Distance and Zero Gauge Coupling in 5d Supergravity}

\author{Ben Heidenreich}
\email{bheidenreich@umass.edu}
\affiliation{Department of Physics, University of Massachusetts, Amherst, MA 01003, USA}

\author{Tom Rudelius}
\email{rudelius@ias.edu}
\affiliation{School of Natural Sciences, Institute for Advanced Study, Princeton, NJ 08540, USA}

\date{\today}

\preprint{ACFI-T20-08}

\begin{abstract}
In this note, we prove that for all five-dimensional supergravities arising from M-theory compactified on a Calabi-Yau threefold, points of vanishing gauge coupling lie at infinite distance in the moduli space. Conversely, any point at infinite distance in the vector multiplet moduli space is a point of vanishing gauge coupling. This agrees with expectations from the Tower/Sublattice Weak Gravity Conjecture, the Swampland Distance Conjecture, and the Emergence Proposal.
\end{abstract}


\maketitle

\section{Introduction}

Quantum gravities with sufficient supersymmetry tend to have (i) gauge fields and (ii) continuous moduli spaces of vacua controlled by expectation values of scalar fields \cite{Ooguri:2006in}. Recently, much work has gone into understanding the universal behavior that arises in weakly-coupled limits of these gauge theories and in asymptotic, infinite distance limits of these moduli spaces. This had led to a number of ``swampland conjectures'' that attempt to describe this universal behavior, along with other general features of quantum gravities. Among them are the Tower/Sublattice Weak Gravity Conjecture (T/sLWGC) \cite{ArkaniHamed:2006dz, Heidenreich:2015nta, Heidenreich:2016aqi, Andriolo:2018lvp}, the Swampland Distance Conjecture (SDC) \cite{Ooguri:2006in}, and the Emergence Proposal (EP) \cite{Harlow:2015lma, Heidenreich:2017sim, Grimm:2018ohb, Heidenreich:2018kpg}.

The T/sLWGC requires an infinite tower of particles to become massless at a point of vanishing gauge coupling, whereas the SDC requires an infinite tower of particles to become massless at an infinite distance point in moduli space. In turn, the EP suggests that a vanishing gauge coupling should emerge as an infinite tower of charged particles become light \cite{Heidenreich:2017sim}, and infinite scalar field distance should emerge as an infinite tower of particles become light \cite{Grimm:2018ohb, Heidenreich:2018kpg}.

Thus, a supergravity theory satisfying the T/sLWGC will have an infinite tower of massless charged particles in the limit of vanishing gauge coupling, which by the EP should generate an infinite distance in scalar field space. Conversely, a theory satisfying the SDC will have an infinite tower of massless particles in the limit of infinite distance, which by the EP should generate a vanishing gauge coupling (assuming the particles are charged).

In this note, we will see that these expectations are borne out in 5d supergravity theories arising from M-theory compactifications on Calabi-Yau threefolds: points of vanishing gauge coupling are at infinite distance, and points at infinite distance in vector multiplet moduli space have vanishing gauge coupling.\footnote{These points of zero gauge coupling and infinite distance are associated geometrically with curves or divisors of the Calabi-Yau going to infinite size \cite{Lee:2019wij}.} These results complement similar results in 4d $\mathcal{N}=2$ supergravity theories arising from type II compactifications on Calabi-Yau threefolds \cite{Grimm:2018ohb, Gendler:2020dfp} and 6d supergravity theories arising from F-theory compactifications on elliptically-fibered Calabi-Yau threefolds \cite{Lee:2018urn, Lee:2018spm, Lee:2019wij}.

The remainder of the note is structured as follows. We begin with a brief review of relevant properties of Calabi-Yau threefolds and 5d supergravity, and we derive a related positivity lemma. We then prove that points of zero gauge coupling lie at infinite distance in the moduli space as well as the converse statement.

\section{The K\"ahler Cone of a Calabi-Yau Threefold }

A Calabi-Yau threefold $X$ is equipped with a (1,1)-form $J$ known as the \emph{K\"ahler form}, which takes values inside a strongly convex polyhedral cone whose interior is the \emph{K\"ahler cone}, $\mathcal{K}(X)$. Within the K\"ahler cone, $J$ can be expressed as a positive linear combination,
\begin{equation}
J = \sum_i \omega_i t^i,~~ t^i > 0,
\end{equation}
where each $\omega_i, i = 1,...,N$ is called a generator of the K\"ahler cone. If the number of generators $N$ is equal to the dimension $h^{1,1}(X)$ of the K\"ahler cone, the cone is \emph{simplicial}. If $N > h^{1,1}(X)$, the cone is \emph{nonsimplicial}.

Any $(1,1)$-form $\lambda$ in the closure of the K\"ahler cone is called \emph{nef}.
Nef $(1,1)$-forms have the property that their triple product is non-negative: $\lambda_1 \cdot \lambda_2 \cdot \lambda_3 \geq 0$ for $\lambda_i$ nef. In particular, this implies that the triple intersection of any three generators of the K\"ahler cone is non-negative.

If the K\"ahler cone is simplicial, we can take the generators $\omega_I$ of the K\"ahler cone to be a basis of $h^{1,1}(X)$. Thus, we can write $J = \sum_I \omega_I Y^I$, with $Y^I > 0$, and the triple intersection numbers are all non-negative:
\begin{align}
C_{IJK} := \omega_I \cdot \omega_J \cdot \omega_K \geq 0, ~~~ I,J,K =1 ,...,h^{1,1}(X).
\label{eq:triple}
\end{align}
Such a basis does not exist when the K\"ahler cone is nonsimplicial \cite{Rudelius:2014wla}. One can choose a subset of the generators as a basis for $h^{1,1}(X)$, but then one cannot express $J$ as a positive linear combination of these particular generators over the entirety of the K\"ahler cone.

In addition, the K\"ahler form $J$ is often able to cross certain codimension-1 boundaries of the K\"ahler cone $\mathcal{K}(X)$ into the K\"ahler cone $\mathcal{K}(\tilde{X})$ of a birationally-equivalent Calabi-Yau $\tilde{X}$, which is related to $X$ by a flop transition. The union of the K\"ahler cones of all of these birationally-equivalent Calabi-Yau threefolds is called the \emph{extended K\"ahler cone} of $X$, which we denote $\mathcal{K}_\cup(X)$. Within a given K\"ahler cone $\mathcal{K}(X)$, $J$ may be written as $J  = \sum_I \omega_I Y^I$, with $Y^I > 0$ and $C_{IJK} \geq 0$, but this positivity will cease to hold as $J$ passes through the boundary between the two K\"ahler cones, as some $Y^I$ switches from positive to negative.

These subtleties are avoided if $J$ is contained in a particular, simplicial subcone of a particular K\"ahler cone $\mathcal{K}(X)$. In particular, consider generic paths in the extended K\"ahler cone of some Calabi-Yau $X$, 
\begin{align}
\gamma: [0,1] \rightarrow \mathcal{K}_\cup(X) \,,~~~x \mapsto J(x),
\end{align}
where $J(x)$ approaches either an asymptotic boundary or a point of zero gauge coupling as $x \to 0$. If $x_0 >0$ exists such that for $0\leq x\leq x_0$, $J(x)$ in contained within the closure of a particular simplicial subcone of a single K\"ahler cone $\mathcal{K}(X)$ with generators $\{\omega_I\}$, then we may write 
\begin{align}
J(x) =\! \sum_{I=1}^{h^{1,1}(X)} \omega_I Y^I(x) , ~~~Y^I(x) \geq 0~ \text{ for all } x \in [0,x_0] .
\label{eq:Jpath}
\end{align}
Notably, the K\"ahler moduli $Y^I$ depend on $x$, but the basis $(1,1)$-forms $\omega_I$ do not. In this basis, $C_{IJK} \geq 0$ and $Y^I(x) \geq 0$ for all $I,J,K$. We will use this positivity repeatedly in what follows.

\section{5d Supergravity}

Many features of a 5d supergravity are captured by its \emph{prepotential}, a cubic homogeneous polynomial:
\begin{equation}
\mathcal{F} \df \frac{1}{6} C_{I J K} Y^I Y^J Y^K := 1.
\end{equation}
In an M-theory compactification to 5d on a Calabi-Yau threefold $X$, indices $I,J,K$ run from $1$ to $h^{1,1}(X)$, the constants $C_{IJK}$ are the triple intersection numbers of (\ref{eq:triple}), and the moduli $Y^I$ are volumes of calibrated 2-cycles, see (\ref{eq:Jpath}). The constraint $\calF:=1$ follows from the fact that the overall volume of the Calabi-Yau is not a vector multiplet modulus in 5d, so the vector multiplet moduli space has dimension $h^{1,1}(X)-1$, and may be thought of geometrically as the $\mathcal{F}=1$ slice of the extended K\"ahler cone.\footnote{In 4d $\mathcal{N}=2$ compactifications of type IIA string theory, there may be additional non-geometric phases in the vector multiplet moduli space, but in 5d these phases are absent \cite{Witten:1996qb}.} At a generic point in moduli space, the gauge group is $U(1)^{h^{1,1}(X)}$, and the gauge kinetic matrix is given by\footnote{See, e.g., \cite{Bergshoeff:2004kh, Lauria:2020rhc}. We set $2 \kappa_5^2 = 1$, with $Y^I_{\text{(here)}} = \sqrt{3} h^I_{\text{(there)}}$ and $C_{I J K}^{\text{(here)}} = \frac{2}{\sqrt{3}} \mathcal{C}_{I J K}^{\text{(there)}}$.}
\begin{equation}
  a_{I J} =\mathcal{F}_I \mathcal{F}_J -\mathcal{F}_{I J} ,
  \label{eq:gaugekinetic}
\end{equation}
with
\begin{equation}
   \mathcal{F}_I \df \frac{1}{2} C_{I J K} Y^J Y^K, 
   \qquad \mathcal{F}_{I
   J} \df C_{I J K} Y^K.
   \label{eq:prepottrip}
   \end{equation}
The eigenvalues of the gauge kinetic matrix correspond to the inverse-squares of gauge couplings, $\lambda_I \sim 1/g_I^2$. Thus, the eigenvalues of $a_{IJ}$ are positive-semidefinite everywhere in moduli space and positive-definite at a generic points, and an eigenvalue of $a_{IJ}$ blows up precisely when a gauge coupling vanishes.

The metric on moduli space is given by the pullback of the gauge kinetic matrix to the $\calF = 1$ slice of the extended K\"ahler cone,
\begin{equation}
  \left( \frac{d s}{d x} \right)^2 = a_{I J}  \frac{d Y^I}{d x} \frac{d Y^J}{d
  x} = \left( \mathcal{F}_I \frac{d Y^I}{d x} \right)^2 -\mathcal{F}_{I J} 
  \frac{d Y^I}{d x}  \frac{d Y^J}{d x} .
  \label{metric}
\end{equation}

To show that points of zero gauge coupling lie at infinite distance in moduli space, we consider a path in vector multiplet moduli space that approaches a point of zero gauging coupling. As discussed previously, we assume the path in question lies inside a
fixed simplicial subcone of a fixed K{\"a}hler cone as we approach zero gauge
coupling, described by
\begin{equation}
  \forall I, \quad Y^I \geq 0, \label{eqn:Ypositive}
\end{equation}
in an appropriate basis. The intersection numbers are
non-negative in this basis:
\begin{equation}
  \forall I, J, K, \quad C_{I J K} \geq 0 . \label{eqn:Cpositive}
\end{equation}
The ability to choose a basis where~\eqref{eqn:Ypositive} and~\eqref{eqn:Cpositive} hold as zero gauge coupling is approached amounts to
a regularity condition on the path: we assume it does not meander back and forth indefinitely between different K\"ahler cones $\mathcal{K}(X)$, $\mathcal{K}(\tilde X)$ within the extended K\"ahler cone, nor does it endlessly meander in and out of the simplicial subcone of $\mathcal{K}(X)$ generated by the $\omega_I$. This assumption is justified when seeking distance-minimizing paths, as such meandering will only serve to increase the path length. 

Per (\ref{eq:gaugekinetic}), some component of $Y^I$ must go to infinity for an eigenvalue of
$a_{I J}$ to blow up. Let $Y^I = Y^I (x)$ be an arbitrary parameterization of the path, with $x = 0$ the point at which a gauge coupling goes to
zero. We assume a Laurent expansion of the form
\begin{equation}
  Y^I (x) = \sum_{n = - N}^{\infty} x^n Y^I_n
\end{equation}
for some $N > 0$, where the path lies entirely within the cone $Y^I
\geq 0$ for a finite interval $0 < x < x_0$.\footnote{This ansatz is more general than it appears. For instance, by appropriately redefining $x$, such a Laurent expansion exists for any path specified by algebraic functions $Y^I(x)$. (Algebraic functions occur naturally in this context due to the polynomial nature of the prepotential.)}

To show that zero gauge coupling lies at infinite distance, we proceed by contradiction: assuming the path length to be finite, we derive $Y^I_n = 0$ for all $n<0$, and therefore $Y^I(x)$ is finite as $x\to 0$, so $a_{I J}$ is also finite.

\section{A positivity lemma}

We first derive a basic consequence of~\eqref{eqn:Ypositive}, \eqref{eqn:Cpositive}, which we use repeatedly in what follows.
 While clearly 
\begin{equation}
  C_{I J K} Y^I Y^J Y^K \geq 0,
\end{equation}
in general there is no constraint of the form $C_{I J K} Y^I_m Y^J_n
Y^K_p \geq 0$ on the Laurent coefficients. The following lemma establishes when such a constraint holds:
\begin{align}
  \text{If} \quad C_{I J K} Y^I_{q} Y^J_{r} Y^K_{s} = 0 &\quad
  \text{for all} \quad (q,r,s) < (m, n, p) \nonumber \\
  \text{then} \quad C_{I J K}&
  Y^I_m Y^J_n Y^K_p \geq 0 \,. \label{eqn:poslemma}
\end{align}
Here we define a partial order on tuples: 
\begin{equation}
(i, j, k) \leq (i',
j', k') \quad \text{if}\quad i \leq i', j \leq j', k \leq k',  \\
\end{equation}
and $(i, j, k) < (i', j', k')$ indicates distinct ordered tuples. 

To prove~\eqref{eqn:poslemma}, let $S_n$ be the set of indices $I$ for which the Laurent expansion of $Y^I(x)$ has leading term $x^{n}$ (i.e., for which $Y^I_n \ne 0$ and $Y^I_{m<n} = 0$). This divides the indices $I$ into disjoint sets $S_n, n = - N, \ldots,
\infty$. Writing an index $I$ restricted to
lie within $S_n$ as $I_n$, \eqref{eqn:Ypositive}
implies
\begin{equation}
  Y_n^{I_n} > 0, \label{eqn:YLaurentPositive}
\end{equation}
as the leading term cannot be negative as $x\rightarrow 0$. Thus:
\begin{equation}
  C_{I J K} Y^I_m Y^J_n Y^K_p = \sum_{(q, r,s) \leq (m,
  n, p)} \!\!\!\!\!\!\!\! C_{I_{q} J_{r} K_{s}}  Y^{I_{q}}_m Y^{J_{r}}_n
  Y^{K_{s}}_p.~
\end{equation} ~

Next, we show inductively that $C_{I J K} Y^I_{q} Y^J_{r}
Y^K_{s} = 0$ for $(q,r,s) \leq (m, n, p)$ if and only if
$C_{I_{q} J_{r} K_{s}} = 0$ for $(q,r,s) \leq (m, n, p)$. Clearly
\begin{equation}
  C_{I J K} Y^I_{- N} Y^J_{- N} Y^K_{- N} = C_{I_{- N} J_{- N} K_{- N}}
  Y^{I_{- N}}_{- N} Y^{J_{- N}}_{- N} Y^{K_{- N}}_{- N},
\end{equation}
which vanishes if and only if $C_{I_{- N} J_{- N} K_{- N}} = 0$ because
$Y^{I_{- N}}_{- N} > 0$. Now suppose that $C_{I J K}
Y^I_{q} Y^J_{r} Y^K_{s} = 0$ and $C_{I_{q} J_{r} K_{s}} = 0$ for
all $(q, r, s) \leq (m, n, p)$ for some particular $m, n, p$. We find
\begin{equation}
  C_{I J K} Y^I_{m+1} Y^J_n Y^K_p = C_{I_{m + 1} J_n K_p} Y^{I_{m
  + 1}}_{m+1} Y^{J_n}_n Y^{K_p}_p,
\end{equation}
which vanishes if and only if $C_{I_{m + 1} J_n K_p} = 0$. The same conclusion follows when
incrementing $n$ or $p$, completing the proof by induction.

Thus, if $C_{I J K} Y^I_{q} Y^J_{r} Y^K_{s} = 0$
for all $(q, r, s) < (m, n, p)$ then $C_{I_{q} J_{r} K_{s}} = 0$ for all $(q,
r, s) < (m, n, p)$, implying
\begin{equation}
  C_{I J K} Y^I_m Y^J_n Y^K_p = C_{I_m J_n K_p} Y^{I_m}_m
  Y^{J_n}_n Y^{K_p}_p \,.
\end{equation}
This is non-negative per~\eqref{eqn:Cpositive}, \eqref{eqn:YLaurentPositive}, so~\eqref{eqn:poslemma} is proven.

\section{Proof of infinite distance}

Consider the constraint
\begin{equation}
  \frac{1}{6} C_{I J K} Y^I Y^J Y^K = 1,
\end{equation}
and expand the left-hand side in negative powers of $x$. We show that
\begin{equation}
  C_{I J K} Y^I_m Y^J_n Y^K_p = 0 \qquad \mathrm{for} \qquad m + n +
  p < 0 \,. \label{eqn:NegTotal}
\end{equation}
In particular, the leading power of $x$ gives
\begin{equation}
  \frac{1}{6} C_{I J K} Y^I_{- N} Y^J_{- N} Y^K_{- N} x^{- 3 N} = 0,
\end{equation}
therefore $C_{I J K} Y^I_{- N} Y^J_{- N} Y^K_{- N} = 0$. Next, suppose that $C_{I J K} Y^I_m Y^J_n Y^K_p =
0$ for all $m + n + p < - M$, where $M > 0$. The leading power of $x$ now gives
\begin{equation}
  \frac{1}{6}  \sum_{\substack{
    m, n, p\\
    m + n + p = - M}} C_{I J K} Y^I_m Y^J_n Y^K_p x^{- M} = 0 .
\end{equation}
The lemma~\eqref{eqn:poslemma} implies that each term in the sum is non-negative, hence each term
vanishes individually, proving~\eqref{eqn:NegTotal} by induction.

The path length is finite if and only if the Laurent expansion of
$\left( \frac{d s}{d x} \right)^2$ at $x = 0$ has no $1 / x^2$ (log divergent) or more singular (power-law divergent) term.
From (\ref{metric}), we obtain 
\begin{widetext}
\begin{align}
  \left( \frac{d s}{d x} \right)^2 &= \frac{1}{x^2}  \left( \frac{1}{2} 
  \sum_{m, n, p} m C_{I J K} Y_m^I Y_n^J Y_p^K x^{m + n + p}
  \right)^2  - \frac{1}{x^2} \sum_{m, n, p} m n C_{I J K} Y^I_m Y^J_n
  Y^K_p x^{m + n + p}, \nonumber \\
  &= \frac{1}{x^2}  \left( \frac{1}{6}  \sum_{m, n, p}(m + n + p) C_{I J K}
  Y_m^I Y_n^J Y_p^K x^{m + n + p} \right)^2  - \frac{1}{3 x^2} \sum_{m, n, p}(mn+mp+np)C_{I J K} Y^I_m
  Y^J_n Y^K_p x^{m + n + p} .
\end{align}
\end{widetext}
Per~\eqref{eqn:NegTotal}, the most singular power that can occur
is $1 / x^2$. This term takes the form:
\begin{equation}
  \left( \frac{d s}{d x} \right)^2_{- 2} = \frac{1}{3}
  \sum_{m, n} (m^2 + m n + n^2) C_{I J K} Y^I_m Y^J_n Y^K_{- m -
  n} .
\end{equation}
Each summand is non-negative by~\eqref{eqn:poslemma}, therefore
\begin{equation}
  C_{I J K} Y^I_m Y^J_n Y^K_{- m - n} = 0, \qquad (m, n) \neq (0,
  0), \label{eqn:ZeroTotal}
\end{equation}
to have a finite path length.

We now make use of the assumption that the gauge kinetic matrix is
positive-definite. Consider
\begin{equation}
  V^I_t \df \sum_n t_n Y_n^I x^n
\end{equation}
for arbitrary coefficients $t_n$. We have
\begin{align}
  V^I_t\! V^J_t\! a_{I J} &= \biggl[ \frac{1}{2} \!\! \sum_{m, n, p} \! t_m
  C_{I J K} Y^I_m Y^J_n Y^K_p x^{m+n+p} \biggr]^2  \nonumber \\  
  &\mathrel{\phantom{=}}-\!\!\sum_{m,
  n, p} \!\! t_m t_n C_{I J K} Y^I_m Y^J_n Y^K_p x^{m+n+p}\ge 0  \label{eqn:non-neg-cons}
\end{align}
for all $0 < x < x_0$ and arbitrary $t_p$.

Using~\eqref{eqn:non-neg-cons}, we show inductively that
\begin{equation}
  C_{I J K} Y_m^I Y_n^J Y_p^K = 0 \quad \text{when $m < 0$}, \label{eqn:NoNeg}
\end{equation}
for all $n, p$. To do so, note that this statement is true when $m + n + p
\leq 0$ per~\eqref{eqn:NegTotal}, \eqref{eqn:ZeroTotal}. Now suppose that it is true for $m + n + p < M$,
$M > 0$, i.e.,
\begin{align}
  C_{I J K} Y_m^I Y_n^J Y_p^K &= 0, \quad m < 0, \quad m + n +
  p < M,  \\
  C_{I J K} Y_m^I Y_n^J Y_p^K &\geq 0, \quad m < 0, \quad
  m + n + p = M,  \label{eqn:nonNegM}
\end{align}
where~\eqref{eqn:nonNegM} is a consequence of~\eqref{eqn:poslemma}. We choose $t_p = 0$
for $0 \leq p \leq \lfloor M / 2 \rfloor$ and $t_p > 0$ for $p<0$ or $p>\lfloor M / 2 \rfloor$.
This ensures that
\begin{align}
  \frac{1}{2}\!\!  \sum_{m, n, p}\! t_m C_{I J K} Y^I_m Y^J_n Y^K_p x^{m+n+p} &= O(x^M) \!+\! O(x^{\lfloor\! \frac{M}{2}\! \rfloor +
  1}), \nonumber \\
  \!\sum_{m, n, p}\!\! t_m t_n C_{I J K} Y^I_m Y^J_n Y^K_p x^{m+n+p}
  &= O(x^M) \!+\! O(x^{2 \lfloor\! \frac{M}{2}\! \rfloor + 2}),  
\end{align}
where the first term on the right-hand side is the leading contribution from
terms with either $m, n$ or $p$ negative, and the second is the leading
contribution from terms with $m, n$ and $p$ all non-negative. Thus, the
leading contribution to $V^I_t V^J_t a_{I J}$ is at order $x^M$, only the
second term in (\ref{eqn:non-neg-cons}) contributes at this order, and only the
terms in the sum with either $m, n$ or $p$ negative contribute. Thus,
\begin{align}
  V^I_t V^J_t a_{I J} = - \sum_{m, n} t_m t_n C_{I J K} Y^I_m Y^J_n
  Y^K_{M - m - n} x^M \nonumber \\
 +O(x^{M + 1}) .
\end{align}
Each summand is non-positive per~\eqref{eqn:nonNegM}, so each must vanish individually. Symmetrizing:
\begin{equation}
  (t_m t_n + t_m t_p + t_n t_p) C_{I J K} Y^I_m Y^J_n Y^K_p = 0,
  ~~ m + n + p = M .
\end{equation}
Given $m < 0$, the constraint $m + n + p = M$ implies that either $n >
\lfloor M / 2 \rfloor$ or $p > \lfloor M / 2 \rfloor$, hence either $t_m t_n > 0$ or $t_m t_p > 0$, implying $t_m t_n + t_m t_p + t_n t_p > 0$. Thus,
\begin{equation}
  C_{I J K} Y^I_m Y^J_n Y^K_p = 0, ~~ m + n + p = M, ~~ m
  < 0,
\end{equation}
which establishes~\eqref{eqn:NoNeg} by induction.

As a corollary, using~\eqref{eq:gaugekinetic}, \eqref{eq:prepottrip} we obtain 
\begin{equation}
  a_{I J} Y^I_n Y^J_n = 0, \qquad n < 0 .
\end{equation}
Since $a_{I J}$ is assumed positive-definite along the path,
this implies $Y^I_n = 0$ for $n < 0$, hence the Laurent expansion is
actually a Taylor expansion, and $Y^I$ remains finite along the path. This
implies that $a_{I J}$ also remains finite, and in particular we cannot reach
a zero-coupling point along a finite-length path. This establishes our desired result: points of vanishing gauge coupling lie at infinite distance in moduli space.

\section{Proof of Zero Gauge Coupling}

We now show the converse: points at infinite distance in vector multiplet moduli space are necessarily points of vanishing gauge coupling. An analogous result in 4d $\mathcal{N}=2$ supergravity theories has been shown in \cite{Gendler:2020dfp} using asymptotic Hodge theory \cite{schmid, CKS, CattaniKaplan, Kerr2017}, but here we will establish the 5d result using only the cubic nature of the prepotential and the positivity conditions discussed previously.

It will prove useful to extend our discussion away from the $\calF = 1$ slice of the K\"ahler cone, instead letting the $Y^I$ be homogenous coordinates invariant under rescaling $Y^I \rightarrow \lambda Y^I$.
$a_{IJ}$ extends straightforwardly to the space of homogenous coordinates,
\begin{equation}
a_{IJ} = \frac{\calF_I \calF_J}{\mathcal{F}^{4/3}} - \frac{\calF_{IJ}}{\calF^{1/3}},
\label{a}
\end{equation}
Extending the metric requires a little more care. We define projected coordinates
\begin{equation}
\hat{Y}^I = \frac{Y^I}{\calF^{1/3}},
\label{hatted}
\end{equation} 
which necessarily satisfy $C_{IJK} \hat{Y}^I \hat{Y}^J \hat{Y}^K = 6$.
We then have the metric 
\begin{equation}
ds^2 = a_{\hat{K}\hat{L}}(\hat{Y}) d\hat{Y}^I d\hat{Y}^J = a_{\hat{K}\hat{L}} \frac{\partial \hat{Y}^{\hat{K}}}{\partial Y^I} \frac{\partial \hat{Y}^{\hat{L}} }{\partial Y^J}  dY^I dY^J,
\end{equation}
on the space of homogenous coordinates $Y^I$, where $a_{\hat{K}\hat{L}}$ is given by (\ref{eq:gaugekinetic}). Using (\ref{hatted}), this can be written as
\begin{equation}
d s^2 = g_{IJ} dY^I dY^J \,, \qquad
g_{IJ} = \frac{2}{3} \frac{\calF_I \calF_J}{\mathcal{F}^2} - \frac{\calF_{IJ}}{\calF} .
\label{hgmet}
\end{equation}
This metric is positive-semidefinite: all eigenvalues are positive inside the K\"ahler cone except for the null eigenvalue corresponding to rescaling $Y^I \rightarrow \lambda Y^I$.

Next, suppose $Y^I_0$ is a point at infinite distance in moduli space. We will argue that $Y_0^I$ is also a point of vanishing gauge coupling. As before, consider a path with endpoint $Y^I(x=0) = Y_0^I$ that is contained entirely within a region $Y^I \geq 0$ in a basis where $C_{IJK} \geq 0$. This implies that $\calF_{IJ}$, $\calF_I$, and $\calF$ are non-negative for all $I$, $J$. 

By homogeneous rescaling, we can ensure that each $Y^I$ remains finite in the limit $Y^I \rightarrow Y_0^I$, and at least one $Y^L$ remains nonzero. This coordinate choice implies that $\dot Y^I$ also remains finite for all $I$, so infinite distance requires at least one eigenvalue of $g_{IJ}$ to diverge. Since each $Y^I$ is finite, each $\calF_I$ is also finite, which by (\ref{hgmet}) means $\calF$ must vanish at $Y_0^I$.

Next, we show that in our chosen coordinate system there exists $K$ such that $\calF_{K}^2/\calF$ is nonzero at the point $Y_0^I$. Since $Y_0^L$ is nonzero, there exists at least one pair of indices $K$, $M$ such that $\calF_{KM} = C_{KML} Y^L$ remains nonzero in the limit $Y^I \rightarrow Y_0^I$; otherwise, $a_{I J} Y_0^I Y_0^J = 0$ everywhere in moduli space, contradicting positive-definiteness of $a_{IJ}$ in the interior of the K\"ahler cone. If $K=M$, then $g_{KK} \geq 0$ (required by positive-semidefiniteness of $g_{IJ}$) implies $(\calF_K)^2 /\calF$ is nonzero at $Y_0^I$. If $K \neq M$, then consider
\begin{align}\calF^2 g_{KK} g_{MM} \geq \calF^2 |g_{KM}|^2,
\end{align}
which is required by positive-semidefiniteness of $g_{IJ}$. 
 If the right-hand side of this inequality vanishes at $Y_0^I$, then because $\calF_{KM}$ is nonzero, $\calF_K \calF_M / \calF$ must also be nonzero.
Alternatively, if the right-hand side is nonzero, then the left-hand side must also be nonzero. $g_{KK}$ and $g_{MM}$ must be non-negative, so from (\ref{hgmet}) and the fact that $\calF_{IJ}$ is non-negative in our coordinate system for all $I$, $J$, we again conclude that $\calF_K \calF_M / \calF$ is nonzero. This in turn implies that either $\calF_K^2/\calF$ or $\calF_M^2/\calF$ is nonzero, so without loss of generality we may assume $(\calF_{K})^2/\calF$ is nonzero at $Y_0^I$.

Since $(\calF_{K})^2/\calF$ is nonzero and $\calF \rightarrow 0$ as $Y^I \rightarrow Y_0^I$, $(\calF_{K})^2/\calF^{4/3}$ diverges in this limit. From (\ref{a}), this implies $a_{KK}$ diverges in the limit unless $\calF_{KK} /\calF^{1/3}= (\calF_K)^2/\calF^{4/3} + \mathcal{O}(1)$ terms, but in the latter case $g_{KK}$ is negative due to the factor of $2/3$ in the first term of (\ref{hgmet}) relative to (\ref{a}), contradicting semidefiniteness of $g_{IJ}$. We conclude that $a_{KK}$ is infinite at $Y_0^I$, so $Y_0^I$ is indeed a point of vanishing gauge coupling.

Note that our proofs relied heavily on the positivity conditions \eqref{eqn:Ypositive} and \eqref{eqn:Cpositive}, which are ensured by the properties of Calabi-Yau geometry discussed previously. We have not found any counterexamples to the statements we have proven by relaxing these conditions. It would be interesting to find such counterexamples or to develop proofs that do not rely on $C_{IJK}$ positivity.

\begin{acknowledgments}
\vspace*{1cm}
{\bf Acknowledgments.}  We thank Sergio Cecotti, Matthew Reece, Nathan Seiberg, Cumrun Vafa, Irene Valenzuela, Timo Weigand, and Edward Witten for useful discussions. We thank Matthew Reece for comments on a draft. BH was supported by National Science Foundation grant PHY-1914934 during the final stages of this work, and by Perimeter Institute for
Theoretical Physics during its inception. Research at Perimeter Institute is supported
by the Government of Canada through the Department of Innovation, Science and Economic Development, and by the Province of Ontario through the Ministry of Research,
Innovation and Science.
TR was supported by the Roger Dashen Membership and by NSF grant PHY-1911298.
\end{acknowledgments}

\bibliography{refs}

\end{document}